\documentclass[preprint,aps,showpacs]{revtex4}
\usepackage[dvips]{graphicx}
\newcommand{\bT} {\mbox{\boldmath$T$}}
\newtheorem{guess}{Conjecture}
\newtheorem{lem}{Lemma}
\newtheorem{theor}{Theorem}

\begin{document}

\title{Proof of the generalized Lieb-Wehrl conjecture for
integer indices larger than one}
\author{Ayumu Sugita\thanks{sugita@yukawa.kyoto-u.ac.jp}}

\affiliation{Yukawa Institute for Theoretical Physics,\\
Kitashirakawa-Oiwakecho, Kyoto, Kyoto 606-8502, Japan}

\date{\today}

\begin{abstract}
Gnutzmann and \.{Z}yczkowski have proposed the R\'{e}nyi-Wehrl 
entropy as a generalization of the Wehrl entropy,
and conjectured that its minimum is obtained for
coherent states. We prove this conjecture for the
R\'{e}nyi index $q=2,3,...$ in the cases of compact
semisimple Lie groups. 
A general formula for the minimum value is given.
\end{abstract}

\maketitle

\section{Introduction}

The Husimi function \cite{husimi} is a useful tool to investigate 
dynamical properties
of quantum systems. 
Wehrl proposed a classical
entropy \cite{wehrl} using the Husimi function and 
conjectured that the minimum of the entropy is obtained
for coherent states. His conjecture was soon proved by Lieb
\cite{lieb}. However, Lieb's analogous conjecture 
for spin coherent state in the same paper has remained unproved 
for more than 20 years.

The Husimi function for the density operator $\hat{\rho}$ is defined as
\begin{eqnarray}
\rho (\zeta) \equiv \langle\zeta |\hat{\rho} |\zeta\rangle
\end{eqnarray}  
where $|\zeta\rangle$ is a coherent state. 
The generalized coherent states defined by Perelomov \cite{perelomov}
can also be used instead of the standard coherent state.
Gnutzmann and \.{Z}yczkowski \cite{gnutzmann} 
have recently generalized the Wehrl entropy by analogy with the well-known 
R\'{e}nyi entropy \cite{renyi}. The R\'{e}nyi-Wehrl entropy
is defined as 
\begin{eqnarray}
S^{(q)}[\hat{\rho}] = \frac{1}{1-q}\log M^{(q)}[\hat{\rho}].
\label{renyi_wehrl}
\end{eqnarray}
Here, $M^{(q)}$ is the moment of the Husimi function
\begin{eqnarray}
M^{(q)}[\hat{\rho}] = c\int d\mu (\zeta)\, \{\rho(\zeta)\}^{q}
\label{def_moment}
\end{eqnarray}
where $d\mu (\zeta)$ is the Haar measure and $c$ is a 
normalization constant.
In the limit $q\rightarrow 1$, it
reproduces the usual
form of the entropy
\begin{eqnarray}
\lim_{q\rightarrow 1}S^{(q)}[\hat{\rho}] = 
S[\hat{\rho}] \equiv
- c\int d\mu (\zeta)\, \rho (\zeta)\log \rho (\zeta). 
\end{eqnarray}

The generalized Lieb-Wehrl conjecture formed in
\cite{gnutzmann} is the following.
\begin{guess}
The minimum value of the R\'{e}nyi-Wehrl entropy
for $q>0$ is obtained for coherent states.
\end{guess}
We prove this conjecture for integer indices $q\ge 2$
in the cases of the coherent states of compact semisimple
Lie groups. 
The case of spin ($SU(2)$) coherent states has already
been proved \cite{gnutzmann, schupp}. In the case of the 
standard coherent state, $q=1$ corresponds to the original
Wehrl's conjecture proved in \cite{lieb}, and 
theorem 3 therein gives the proof for arbitrary $q>1$.

\section{Generalized coherent states}

Before coming to the proof, let us review 
the definition and some properties of
the generalized coherent states \cite{perelomov}. 
Let $G$ be a compact semisimple Lie group. 
The Lie algebra of $G$ can be written in so-called Cartan basis 
$\{H_{j}, E_{\alpha}\}$. An irreducible representation
of $G$ is characterized by the lowest (or the highest) weight.
Let $|-\lambda\rangle$ be the lowest weight state with
the weight $-\lambda$. We denote the irreducible representation space
specified by $|-\lambda\rangle$ as $D_{\lambda}$. 
Coherent states in $D_{\lambda}$ are obtained by the action of the
group $G$ on $|-\lambda\rangle$,
which can be written explicitly as
\footnote{In fact, (\ref{def_coh}) cannot represent the coherent
states orthogonal to $|-\lambda\rangle$. However, it does not matter when
we consider entropies such as (\ref{renyi_wehrl}) 
because the measure of the set
of such states is zero.}
\begin{eqnarray}
|\zeta\rangle = {\cal N}(\zeta)
\exp (\zeta_{\alpha}E_{\alpha})|-\lambda\rangle.
\label{def_coh}
\end{eqnarray}
Here, ${\cal N}(\zeta)$ is the normalization constant
and $\alpha$ runs over all positive roots. 
The expansion of the exponential function becomes
a finite series because $D_{\lambda}$
is finite dimensional. Therefore
\begin{eqnarray}
\tilde{\psi}(\zeta) \equiv 
\frac{\langle\psi |\zeta\rangle}{{\cal N}(\zeta)}
\label{tilde}
\end{eqnarray}
is a polynomial of $\zeta$ for any state $|\psi\rangle$.
We will use this fact later to prove lemma \ref{pure}.

The ``resolution of unity''
\begin{eqnarray}
I_{D_{\lambda}} 
= {\rm dim}D_{\lambda} \int d\mu (\zeta) |\zeta\rangle\langle\zeta |
\label{unity}
\end{eqnarray} 
is valid, and hence the normalization constant
in (\ref{def_moment})
should be taken as
\begin{eqnarray}
c_{D_{\lambda}} = {\rm dim}D_{\lambda},
\end{eqnarray}
in order to satisfy  the normalization condition $M^{(1)}=1$.

\section{Proof of the main result}

Turning now to the proof, 
we first note the following lemma. 
\begin{lem}
If $\hat{\rho}$ minimizes $S^{(q)}[\hat{\rho}]$
for some $q>0$, $\hat{\rho}$ must be a pure state.
\label{pure}
\end{lem} 
This is a generalization of lemma 2 in \cite{lieb},
and can be proved in the same way (see the appendix).
Hereafter we concentrate on pure states.

Let us consider the simplest case $q=2$. We will see later that
the discussion in this case can be easily generalized.
We have to prove that the maximum of the second moment
of the Husimi function
\begin{eqnarray}
M^{(2)}_{|\varphi\rangle} \equiv c_{D_{\lambda}}\int d\mu (\zeta)
|\langle\zeta|\varphi\rangle|^{4}
\label{second_moment}
\end{eqnarray}
is given by coherent states.

Let us rewrite (\ref{second_moment}) as
\begin{eqnarray}
M^{(2)}_{|\varphi\rangle} & = &
\left(\langle\varphi |\otimes\langle\varphi |\right)\,
c_{D_{\lambda}}\int d\mu (\zeta) 
\left(|\zeta\rangle\otimes |\zeta\rangle\right)
\left(\langle\zeta |\otimes\langle\zeta |\right)\,
 \left(|\varphi\rangle \otimes |\varphi\rangle \right).
\end{eqnarray}
The key observation is that the tensor product of 
the coherent states $|\zeta\rangle\otimes |\zeta\rangle$
is a coherent state in the irreducible representation space
$D_{2\lambda}$ whose lowest weight state is 
$|-\lambda\rangle\otimes |-\lambda\rangle$. The same idea
is used in \cite{schupp} for $SU(2)$.
It can be shown explicitly as
\begin{eqnarray}
(H_{j}\otimes I + I\otimes H_{j}) |-\lambda\rangle\otimes
|-\lambda\rangle & = &
-2\lambda_{j}|-\lambda\rangle\otimes |-\lambda\rangle\\
(E_{-\alpha}\otimes I + I\otimes E_{-\alpha}) 
|-\lambda\rangle\otimes |-\lambda\rangle
& = & 0
\end{eqnarray}
\begin{eqnarray}
|\zeta\rangle\otimes |\zeta\rangle
& = & {\cal N}(\zeta)^{2}
\exp [\zeta_{\alpha}(E_{\alpha}\otimes I + I\otimes E_{\alpha})]
|-\lambda\rangle\otimes |-\lambda\rangle
\end{eqnarray} 
where $\alpha$ runs over positive roots.
Therefore $\int d\mu (\zeta) 
\left(|\zeta\rangle\otimes |\zeta\rangle\right)
\left(\langle\zeta |\otimes\langle\zeta |\right)$ is proportional
to the projection operator to $D_{2\lambda}$ in 
$D_{\lambda}\otimes D_{\lambda}$, i.e.
\begin{eqnarray}
c_{D_{\lambda}}\int d\mu (\zeta) 
\left(|\zeta\rangle\otimes |\zeta\rangle\right)
\left(\langle\zeta |\otimes\langle\zeta |\right)
= \frac{{\rm dim}\, D_{\lambda}}{{\rm dim}\, D_{2\lambda}}\, 
P_{D_{2\lambda}}.
\end{eqnarray} 
Here the proportional constant is determined from (\ref{unity}).
Hence
\begin{eqnarray}
M^{(2)}_{|\varphi\rangle} & = &
\frac{{\rm dim}\, D_{\lambda}}{{\rm dim}\, D_{2\lambda}}
\left(\langle\varphi |\otimes\langle\varphi |\right)
P_{D_{2\lambda}}
 \left(|\varphi\rangle \otimes |\varphi\rangle \right)
\end{eqnarray}
and we have
\begin{eqnarray}
|\varphi\rangle\otimes |\varphi\rangle \in D_{2\lambda}\;\; 
\Longleftrightarrow\;\;
M^{(2)}_{|\varphi\rangle}=M^{(2)}_{max} \equiv
\frac{{\rm dim}\, D_{\lambda}}{{\rm dim}\, D_{2\lambda}}.
\end{eqnarray}

Then we should prove
\begin{eqnarray}
|\varphi\rangle : \mbox{coherent state}\;\;
\Longleftrightarrow\;\;
|\varphi\rangle\otimes |\varphi\rangle \in D_{2\lambda}. 
\label{tobeproved}
\end{eqnarray}
LHS $\Rightarrow$ RHS is obvious, because $|\varphi\rangle
\otimes |\varphi\rangle$ is obtained by the action
of $G$ on $|-\lambda\rangle\otimes|-\lambda\rangle$
which belongs to $D_{2\lambda}$ by definition.

Then we prove LHS $\Leftarrow$ RHS.
Let us consider the quadratic Casimir operator $C_{2}$
of the group $G$.
By taking an orthonormal basis $\{T_{a}\}$ of the Lie
algebra, it can be written as
\begin{eqnarray}
C_{2} = \bT^{2} \equiv \sum_{a}T_{a}^{2}.
\end{eqnarray}
In the tensor product space $D_{\lambda}\otimes D_{\lambda}$, 
the quadratic Casimir
operator is written as
\begin{eqnarray}
C_{2}^{(2)} 
\equiv \sum_{a}(T_{a}\otimes 1 + 1\otimes T_{a})^{2}
= \bT^{2}\otimes 1 + 1\otimes \bT^{2}
+ 2\sum_{a}T_{a}\otimes T_{a}.
\end{eqnarray}
Therefore
\begin{eqnarray}
(\langle\varphi |\otimes\langle\varphi |)\,C_{2}^{(2)}\,
(|\varphi\rangle\otimes |\varphi\rangle) = 
2<\bT^{2}> + 2 <\bT>^{2}
\end{eqnarray}
where the symbol $<>$ denotes the expectation value for 
$|\varphi\rangle$. 

Here we note the lemma proved in \cite{delbourgo}:
\begin{lem}[Delbourgo and Fox]
The minimum of the uncertainty $(\Delta T)^{2}\equiv
<\bT^{2}>-<\bT>^{2}$ is obtained for coherent states 
\begin{eqnarray}
(\Delta T)^{2}\, \mbox{is minimum}\;
\Longleftrightarrow\;
|\varphi\rangle : \mbox{coherent state}.
\end{eqnarray}
\end{lem}
Since $C_{2}=\bT^{2}$ is a constant in $D_{\lambda}$,
$<\bT>^{2}$ is maximum when $|\varphi\rangle$
is a coherent state, 
and hence $C_{2}^{(2)}$ takes the
maximum value at $D_{2\lambda}$ because it
contains the tensor product of the coherent states.
Therefore
\begin{eqnarray}
|\varphi\rangle\otimes |\varphi\rangle \in D_{2\lambda}\;
\Longrightarrow\;
\mbox{$<\bT>^{2}$ is maximum}\;
\Longrightarrow
|\varphi\rangle :\mbox{coherent state}.
\end{eqnarray}
This completes the proof of (\ref{tobeproved}).

The generalization to the higher moments is straightforward.
By considering the tensor product space 
$D_{\lambda}^{\otimes q} \equiv
\overbrace{D_{\lambda}\otimes D_{\lambda}\otimes\dots 
\otimes D_{\lambda}}^{q}$,
one can easily show
\begin{eqnarray}
M_{|\varphi\rangle}^{(q)} \equiv
\frac{{\rm dim}\, D_{\lambda}}{{\rm dim}\, D_{q\lambda}}
\left(\langle\varphi|^{\otimes q}\right)P_{D_{q\lambda}}
\left(|\varphi\rangle^{\otimes q}\right)
\end{eqnarray}
and 
\begin{eqnarray}
\left(\langle\varphi |^{\otimes q}\right) C_{2}^{(q)}
\left(|\varphi\rangle^{\otimes q}\right)
 = q <\bT^{2}> + q(q-1)<\bT>^{2}
\end{eqnarray}
in obvious notation.
From these two formulae we can conclude
\begin{eqnarray} 
M^{(q)}_{|\varphi\rangle}=M^{(q)}_{max}
\Longleftrightarrow\;
|\varphi\rangle^{\otimes q}\in D_{q\lambda}\;
\Longleftrightarrow\;
|\varphi\rangle : \mbox{coherent state}.
\end{eqnarray}
Combining this and Lemma \ref{pure},
we obtain the main result of this paper:
\begin{theor}
Let $S^{(q)}$ be the R\'{e}nyi-Wehrl entropy with an
integer index $q\ge 2$ of
a compact semisimple Lie group.
The minimum value of $S^{(q)}$
is obtained for coherent states
\begin{eqnarray}
S^{(q)}[\hat{\rho}]\, \mbox{is minimum}\;\;
\Longleftrightarrow\;\;
\hat{\rho} = |\varphi\rangle\langle\varphi|,\;\;
|\varphi\rangle: \mbox{coherent state}.
\end{eqnarray}
The minimum value $S^{(q)}_{min}$ is given by 
\begin{eqnarray}
S^{(q)}_{min} = \frac{1}{1-q}\log
\left(\frac{{\rm dim}\, D_{\lambda}}{{\rm dim}\, D_{q\lambda}}\right)
\label{formula}
\end{eqnarray}
where $D_{\lambda}$ is the irreducible representation space
in which $S^{(q)}$ is defined.
\end{theor}

\section{Minimum values for some special cases}

Finally, let us examine the general formula for the minima
(\ref{formula}) in some special cases.
Cases (i), (ii), and (iii) in the following can be used to investigate the
dynamics of bosons, fermions and distinguishable particles,
respectively 
\footnote{$U(N)$, instead of $SU(N)$, is
used in \cite{sugita2} for bosons and fermions. 
$U(N)$ is not semisimple. However, $U(N)\sim U(1)\times SU(N)$,
and the Abelian subgroup $U(1)$
corresponds to the total phase which is irrelevant
for the Husimi functions.} \cite{sugita2}. 
\begin{itemize}
\item[(i)] $G=SU(N)$, $D_{\lambda}=[1^{m}]$ (symmetric
product of $m$ fundamental representations)
\begin{eqnarray}
{\rm dim}\, D_{q\lambda} = {\rm dim}\, [1^{qm}]
= \frac{(N+qm-1)!}{(N-1)!\,(qm)!} 
\end{eqnarray}
\begin{eqnarray}
S^{(q)}_{min} = \frac{1}{1-q}\log 
\frac{\Gamma (N+m)\, \Gamma (qm+1)}{\Gamma (N+qm)\, \Gamma (m+1)}.
\label{boson}
\end{eqnarray}
Assuming that (\ref{boson}) holds also for non-integer $q$, 
we obtain
\begin{eqnarray}
\lim_{q\rightarrow 1} S^{(q)}_{min} = m\{\psi (N+m)-\psi (m+1)\}
\end{eqnarray}
where $\psi(z)\equiv\frac{d}{dz}\log\Gamma (z)$ is the digamma function.
This result agrees with the value of $S^{(1)}$ for coherent states 
given in \cite{slomczynski}, except for the shift by $-\ln N$
due to another normalization of the coherent states.

\item[(ii)] $G=SU(N)$,
$D_{\lambda}=[m]$ (anti-symmetric product of $m$ fundamental
representations)
\begin{eqnarray}
{\dim}\, D_{q\lambda} = {\rm dim}\, [m^{q}]
& = & \prod_{j=0}^{m-1}\frac{(N+q-j-1)!\,j!}{(N-j-1)!\,(q+j)!}
\end{eqnarray}
\begin{eqnarray}
S^{(q)}_{min} = \frac{1}{1-q}\sum_{j=0}^{m-1}\log
\frac{\Gamma(N+1-j)\Gamma(q+j+1)}{\Gamma(N+q-j)\Gamma(j+2)}.
\end{eqnarray}
In this case, the minimum value for $q=1$ is expected to be 
\begin{eqnarray}
\lim_{q\rightarrow 1} S^{(q)}_{min} = 
\sum_{j=0}^{m-1}\{\psi(N-j+1)-\psi(j+2)\}.
\end{eqnarray}

\item[(iii)] $G=\overbrace{SU(N)\times \dots\times SU(N)}^{m}$,
$D_{\lambda}=[1]^{\otimes m}$
\begin{eqnarray}
{\rm dim}\, D_{q\lambda} = {\rm dim}\, [1^{q}]^{\otimes m}
= \left(\frac{(N+q-1)!}{(N-1)!\,q!}\right)^{m}
\end{eqnarray}
\begin{eqnarray}
S^{(q)}_{min} = \frac{m}{1-q}\log
\frac{\Gamma(N+1)\Gamma(q+1)}{\Gamma(N+q)}.
\end{eqnarray}
The minimum value for $q=1$ is expected to be
\begin{eqnarray}
\lim_{q\rightarrow 1}S^{(q)}_{min} =
m \{\psi (N+1)-\psi (2)\}.
\end{eqnarray}

\end{itemize}

\begin{acknowledgments}
The author thanks Professor Gnutzmann for letting him know about 
reference \cite{schupp}. He also thanks 
Professor \.{Z}yczkowski for helpful discussions.
\end{acknowledgments}

\appendix

\section{Proof of Lemma \ref{pure}}
Here we prove lemma \ref{pure}: 
{\it if $\hat{\rho}$ minimizes $S^{(q)}[\hat{\rho}]$
for some $q>0$, $\hat{\rho}$ must be a pure state}.
 
We consider three cases (i) $0<q<1$, (ii) $q=1$ and
(iii) $1<q$ separately. $\hat{\rho}$ minimizes
$M^{(q)}$ and $S$ in cases (i) and (ii), respectively, and 
maximizes $M^{(q)}$ in case (iii).
Let us decompose the density operator as 
$\hat{\rho}=\sum_{i}\lambda_{i}\hat{\rho}_{i}$,
where
$\sum_{i}\lambda_{i}=1$ and
$\hat{\rho}_{i}\equiv|\psi_{i}\rangle\langle\psi_{i}|$ 
is a pure state.
Since $m^{(q)}(x)=x^{q}\;(0<q<1)$ and $s(x)=-x\log x$ are concave,
and $m^{(q)}(x)=x^{-q}\;(q>1)$ is convex,
\begin{eqnarray}
M^{(q)}[\hat{\rho}] & \ge & 
\sum_{i}\lambda_{i}M^{(q)}[\hat{\rho}_{i}]\;\;\; (0<q<1)\\
S[\hat{\rho}] & \ge & 
\sum_{i}\lambda_{i}S[\hat{\rho}_{i}]\\
M^{(q)}[\hat{\rho}] & \le & 
\sum_{i}\lambda_{i} M^{(q)}[\hat{\rho}_{i}]\;\;\; (q>1).
\label{concave}
\end{eqnarray} 
The equality must hold in at least one of the three inequalities.
Then 
$\rho_{i}(\zeta)=\rho_{j}(\zeta)$, and hence 
$\left|\tilde{\psi_{i}}(\zeta)\right|=
 \left|\tilde{\psi_{j}}(\zeta)\right|$
in the notation of (\ref{tilde}), almost
everywhere for all $i,j$. Since $\tilde{\psi}_{i}$
is a polynomial, $\tilde{\psi}_{i}=a\tilde{\psi}_{j}$, and hence 
$|\psi_{i}\rangle = a^{*}|\psi_{j}\rangle$,
with $|a|=1$. Therefore $\hat{\rho}_{i}=\hat{\rho}_{j}$ for all $i,j$,
which means $\hat{\rho}$ is a pure state.

\end{document}